\documentclass[conference]{IEEEtran}
\usepackage{1Preamble}

\begin{document}

\title{  How  Crucial Is It for  6G Networks \\
to Be Autonomous?}

    
\author{\IEEEauthorblockN{Nadia Adem, Ahmed Benfaid, Ramy Harib, and Anas Alarabi}
\IEEEauthorblockA{\textit{Department of Electrical and Electronic Engineering} \\
\textit{University of Tripoli, Tripoli, Libya} \\
{  E-mail: \{n.adem,a.benfaid,r.harib,an.alarabi\}@uot.edu.ly}\\ \\
}\\ \\
 
\Large{Invited Paper}
}

\maketitle
\begin{abstract}
 
The sixth generation (6G),  unlike any of the previous generations, is envisioned by 2030 to connect everything. Moreover, in addition to the new use cases 6G is expected to support, it will  need to provide {{a}} superior performance over  5G. The global connectivity,  large network dimensions, users heterogeneity,  extremely low-power consumption,  {high throughput, ultrahigh reliability}, efficient network operation and maintenance, and {low-latency} requirements to be met by future networks  inevitably necessitate   the autonomy of 6G. 
  Intelligence, facilitated mainly by the advancement   of   artificial intelligence (AI) techniques, is a key to achieve  autonomy. 
  In this paper{,} we provide a bird’s-eye view of 6G, its vision, progress, and objectives. Furthermore, we  present some     technologies that would be mainly enabling intelligent globally   connected world. In addition to discussing the role of  AI for future wireless communications, we, unlike any other review papers, provide our original results which give early evidence for the viability of achieving 6G networks autonomy   through leveraging AI advances. 
Furthermore, we, very importantly, identify  6G implementation challenges and  key innovative techniques 
that promise to solve them. 
This  article serves as a starting point for  learners to acquire more  knowledge   about 6G and also  for researchers   to promote
more development to the field.
\end{abstract}
 
\section{Introduction} 
\label{sec: Introd}
Recently, as the fifth generation (5G) standard has been getting finalized, and in order to set unified precise targets and roadmap for the sixth generation (6G) communication networks, researchers from industry and academia have begun to envision and extensively discuss its key values and use cases{~\cite{dang2020should,zhang20196G,david20186g,yastrebova2018future,tariq2020speculative,yaacoub2019key,parssinen2020white}.}
\subsection{6G Vision}
The  6G is depicted, in its first global vision,  as a framework   of different services such as sensing, computing, caching, imaging, highly accurate positioning and mobility, radar  and navigation   integrated with the main communication services the previous generations provide. 
6G will be an autonomous intelligent ecosystem  {that} connects everything through multi-dimensional  networks that provide services in the ground, air, space,  and underwater aiming  to provide quality of not just service but life. Benefiting from smart sensory environments, 6G will be driven by a variety of verticals including factories, automation and transportation,  and healthcare. 
6G is predicted to be motivated by potential new applications  
 for which the currently deployed 5G infrastructure is not expected to support~\cite{saad2020avision}.
A main domain of these applications is the connected robotics and autonomous systems that include self-driving cars, drone delivery systems, and autonomous robotics.  These applications will get the full use of the multi-dimensional network structure, and artificial intelligence  {(AI)} capabilities to be offered.
 Other  domains include extended reality  {(XR)}, blockchain  and wireless brain-computer interaction based applications.  Different requirements for these applications and their corresponding use cases set different trends and research directions towards 6G.
 For example,  autonomous systems applications
    require ultra-reliable, low-latent , and secure communications. 
There {are} a number of applications, e.g. holographic communications,
that are bandwidth intensive and thus require opening up new and wider spectrum such as terahertz  {(THz)} bands~\cite{chaccour2021seven}. On the other hand, other use cases involve communications between  multiple small devices that mainly work on batteries  set the trend for energy sustainability. More details about the  requirements demanded, and technologies promising for realizing the 6G are presented in this article.   
\subsection{ 6G objectives }   
Spectrum and energy efficiency, peak data rate, user data rate,   capacity per unit volume, connectivity density, latency, reliability, and mobility are some of the key performance indicators (KPIs)  evaluating 6G networks. Technical objectives  for these KPIs, set mainly based on  requirements, trends, and applications, are summarized  and compared to their counterparts in 5G as follows:

\begin{itemize}
\item Up to $10$ {Tbit/s} of peak data rate ($1000$  times that of 5G) to be targeted~\cite{zhang20196G}. 

\item A user data rate of $1$ Gbit/s ($10$ times 5G) or higher.  


\item  No more than ${100}$ µs    ($1/10$ 5G)  of latency is allowed.  $10$ times lower latency in certain cases needs to be maintained~\cite{zhang20196G}.

\item   Providing services to $10^7$ devices/km$^2$ resulting in $10$ times  of the 5G connectivity density. 

\item  \textit{Volumetric} spectral and energy  efficiency $100$ times  the \textit{per unit area} efficiency in 5G~\cite{saad2020avision}.

\item Reliability of seven $9$s as opposite to that of five $9$s  in 5G. 

\item Support mobility up to ${1000}$ kmph, as the 6G is foreseen  to support airline systems and high-speed railways. The 5G, however, was intended to support less than half of that.  

\end{itemize}

In  this paper we discuss the key  {innovative} technologies promising to meet some of the aforementioned  objectives. 

\subsection{Paper Organization}
This paper is organized as follows. Section \ref{sec:tech} contains some of 6G  enabling technologies\textquotesingle~basic principles and compelling features{,} and discusses the need of integrating them with other techniques  to get the most out of them. In Section \ref{sec:AI}, the role of AI in maintaining intelligent and hence autonomous 6G network is discussed. The Section, in addition, presents some of our simulated experiments results that demonstrate the effectiveness and superiority of AI{-}enabled networks. In Section  \ref{sec:chalg}, we cover some of   future networks\textquotesingle  challenges,  possible solutions, and related open research directions.  Concluding remarks are given in Section \ref{sec:conc}. 

\section{6G enabling Technologies}
\label{sec:tech}
The capability expansions, performance  improvements, variety of new trends and service classes {that need to be} guaranteed and supported by  6G require the incorporation of  disruptive technologies. In this section we present some of 6G promising technologies. 
 
\subsection{Above 6GHz Communications}
The sub-6 GHz band becomes highly congested due to heavily used frequency resources. It  no longer can  support massive increases in communication capacity. 
Allowing accessibility to higher frequencies and bandwidth,  millimeter wave (mmWave) technology has been emerging to overcome the lack of spectrum issue.   Facilitating the 6G high-rate high-mobility use cases  (e.g.  autonomous vehicles), mmWave is considered to be one of the 6G enabling technologies~\cite{chaccour2021seven}.    
 MmWave communications, ranging from  $24$ GHz to $300$ GHz, {will} be one of the leading candidate systems for future wireless communications~\cite{xiao2017millimeter}. 
MmWave   achieves peak data rates of $10$ Gbit/s or more with full-duplex capability, far exceeding the lower microwave frequency limit of $1$ Gbit/s~\cite{rappaport2017overview}. 

Despite its benefits, nevertheless, atmospheric absorption   highly  attenuate{s} mmWave signals. 
  Using  Friis Law~\cite{rappaport1996wireless}, one can determine an  obtained power, $P_R$, at a receiver  in free space   environments (one may refer to~\cite{heath2016overview} for more generalized environment)
  located at a distance $d$ apart from a  transmitter sending a  signal with wavelength  $\lambda$ and power $P_T$   as described by the following equation.

\begin{equation}
\label{eqn: FSL}
   {P_R} = {P_T}G_T G_R (\frac{\lambda}{4 \pi d})^2, 
\end{equation}
where $G_R~\&~G_T$ are the receiving and transmitting antenna gain respectively.   
We can notice from the equation
that under the same distance and antenna settings as microwave signals below 6 GHz, the free space path loss, i.e. ${{P_T}}/{{P_R}}$, for mmWave signals is much higher. 
 This shows that using mmWave frequencies reduces the transmission distance which is a huge disadvantage in mmWave systems.
Due to their associated signals{\textquotesingle} short wavelengths, nevertheless,  mmWave technology can be integrated with  large-scale multi-element antenna arrays, e.g. super massive (SM) multiple input multiple output {(MIMO)}~\cite{yang20196g}, and   beamforming techniques    
allowing for high antenna gain and highly directional communications hence extending communication distance, enhancing security,  and improving interference immunity.  
Another  major technical challenge  for mmWave is the blockage effect.  
Rending to their abilities in providing  line-of-sight (LoS) links with a high chance,  aerial base stations (BSs) and the resultant potential 6G 3D networks can decrease the impact of the aforementioned issue. Furthermore, fusing mmWave with machine learning (ML) and deep learning (DL) techniques to make data-driven decisions allow  getting the best out of it~\cite{Benfaid2021thesis, benfaid2021adaptsky, khosravi2020learning}.  

Commercial mmWave communications have become a reality with the standardization of 5G,
 however their  potentials would fall short of many new application, such as 3D gaming and XR~\cite{saad2020avision}.
Hundreds of gigabits per second to multiple terabits per second data rates are needed for these applications which will require some other innovative solutions. {This} leads us to THz communication  systems  which  are widely recognized as the next step of wireless communications research. 
The  0.275  to 3 terahertz band spectrum is the main part of the THz band,  which ranges from 0.1 to 10 terahertz, to be used for cellular communications, according to International Telecommunication Union Radio Communication Sector (ITU-R) guidelines~\cite{ITU2015THz}. Such  abundance in spectrum portend to offer extremely high data rates~\cite{chaccour2021seven}. 


{Despite their features, however, there are many open research issues in THz radio communications such as small propagation range due to high propagation and molecular absorption losses, and THz transceiver design that will need to be addressed to make it realizable. To achieve a successful THz deployment and operation in 6G networks, nevertheless, 
pencil beams~\cite{tan2020thz,rappaport2019wireless}, and holographic surfaces~\cite{huang2020holographic} solutions are to be considered for  overcoming THz issues. 
In spite of that, the coexistence of THz systems with sub-6 GHz and mmWaves is necessary, especially in dynamic  environments with  non-line-of-sight links where  lower band  technologies can offer  higher reliability and longer ranges.
 }



\subsection{Non-Orthogonal Multiple Access (NOMA)}
{Traditional orthogonal multiple access (OMA) schemes
may struggle to accommodate the massive number of 6G connections, since  these techniques divide a resource block (RB), time, frequency, or code, between  users   equally without taken into consideration the variations of their channel conditions.   
 Capitalizing opportunistic channel conditions, however, NOMA assigns different amount of power over same RB, where worse channel gain translates into a higher power.  As a result,  {on} the contrary to OMA, NOMA exploits the dimension of  power domain to  utilizes a RB  for serving multiple users simultaneously. 
 NOMA, as a consequence, offers not only higher spectral efficiency and throughput but also  better fairness, low{er} service latency, and higher connectivity density~\cite{ding2017survey}.} 
Thanks   to superposition coding (SC), which  encodes multiple signals with distinct power into a single signal at a transmitter,  and successive interference cancellation (SIC), implemented at a NOMA receiver to distinguish a signal from other signals, which make the NOMA idea possible.

NOMA, nonetheless, has some limitations, such as receiver computational complexity,  BS channel state information (CSI) prior knowledge, and performance {degradation} as   number of users increases. 
Due to these issues, the 3rd Generation Partnership Project (3GPP) has deferred the use of NOMA for next{-}generation networks, where most of these limitations will not be a concern, and given up {on} making NOMA an option for 5G~\cite{3GPP2018noma5GNR,makki2020survey}.   
 { To benefit from the superiority of NOMA  in fairness, latency, throughput, and connections density in the 6G networks{,}    NOMA will {be} expected to be implemented in  merged  with some other technologies like  mmWave, and aerial BSs.  
 Optimizing a network combining these technologies, nonetheless, would be a complex non-convex problem that  is of  a challenge to be handled with traditional mathematical solutions~\cite{Benfaid2021thesis, benfaid2021adaptsky}, however, we will discuss the viability of  managing them autonomously  by using the power of AI.}

\subsection{Unmanned Aerial Vehicle (UAV)}
 Aerial BSs, or UAVs,  are emerging as one  {of} the 6G essentials as they can provide high data rates and global wireless connectivity.  UAVs enjoy  exclusive features  such as ease of deployment, high LoS links probability, and large degree-of-freedom provided through their controlled mobility~\cite{li2018uav}.
 UAVs can be  deployed to offer wireless connectivity in case of  emergencies such as natural disasters,  overcome  last mile issues,   enhance capacity, replace terrestrial BSs, etc. 
 Allowing UAVs to  adaptively  relocate according to environmental changes and users\textquotesingle~demands
 is one of the most important UAV features.  
The goal of  connecting the unconnected to be met by the 6G will be made possible by  aerial BSs  as their deployment are much easier and more economical  than that of their terrestrial counterparts, especially in rural areas. 
Rending to their ability  to connect with low earth orbit (LEO) satellites, CubeSats, and terrestrial BSs, UAVs  are also envisioned to  be one of  {the} main technologies to realize the 6G multi-dimensional network.  Managing UAVs{\textquotesingle} placement and resource allocation simultaneously, however,  is a challenge that has to be addressed  in order to get the most out of their resultant 3D networks.   
AI advances  are promising, nonetheless, in managing the UAV networks autonomously yet efficiently and effectively~\cite{Benfaid2021thesis, benfaid2021adaptsky}. 
 \subsection{Quantum Communications and Computing}
Quantum-assisted communication is a novel field that can be viewed  as one of the cornerstones for future multi-state networks. Rendering to their capabilities in solving problems exponentially faster than their classical counterparts,  
quantum computers (QC) will  bring  {a} new era in  telecommunication. Their working principle is based on essential concepts of quantum mechanics such as superposition, entanglement, and no-cloning theorem~\cite{Gyongyosi2019Survey}. 
Constructing independent copies of quantum information  is impossible, as the no-cloning theorem states. Thus, offering high communication security which should be a key feature of 6G.
A qubit,  quantum analogy of   a classical bit,  represents a two-level quantum system, where each level is called a state. Instead of representing zero or one only, however,  a qubit can be in a superposition, linear combination,  of both. That means upon measurement, a qubit will be found with some probability in the one or  zero sate. 
For n qubits, a quantum computer can work with $2^n$ quantum states simultaneously.  This parallelism makes quantum computers potentially useful in applications that require the processing of big data. Many efforts have been done by the research community to investigate on the involvement of quantum speedups in various communication areas. The authors in~\cite{botsinis2019Quantum} for instance  {have proposed and presented}  a number of quantum search algorithms that are particularly applicable to wireless communications. Quantum computing merged with different AI techniques, e.g. quantum machine learning (QML), has promising potentials in solving challenges in many aspects of 6G networks and beyond~\cite{syed2019Quantum}.    
 \subsection{Blockchain}
Blockchain  is considered as a technology breakthrough in the recent years that is expected to have an important role in achieving 6G objectives~\cite{hewa2020therole,nguyen2020blockchain}. Blockchain, which is simply a database structure,  offers multiple key characteristics that are useful when used with certain applications. Blockchain stores data in blocks called datablocks, each datablock is attached to the previous one forming a chain of blocks. This chain is replicated and stored across different nodes that form a network. Once a data update request is initiated, it is broadcasted to all nodes where they use their available computational power to verify that it follows some pre-defined rules, once verified, the datablock is linked to the previous one in the chain. 
The described structure along with the operation method enable the blockchain technology to be used in applications that require high level of security, where changing one datablock in a certain node can be easily detected once the other nodes cross-reference each other. 
Other characteristics provided by blockchain-based systems are their scalability, where they can be used for 6G to overcome scaling limitations of centralized conventional networks. Transparency is also a key feature of blockchain that can be used for dynamic resource  management across different small scale operators since the records can be set to be accessed with equal rights without the need for a third party. The distributed  architecture of blockchain is also beneficial in facilitating decentralized systems and thus eliminating single point failures of previous generations systems.  
Consequently, blockchain was proposed to be used in different areas of 6G such as blockchain-based resource management frameworks~\cite{hao2020}, and blockchain-enabled architectures for UAV networks to ensure security and privacy with enhanced network performance~\cite{gupta2021}.\\
 
The aforementioned technologies promise in meeting some of the 6G objectives and enabling  corresponding features  including throughput, latency, connectivity, security, etc.  Other technologies, whether they are  evolving from the 5G or exclusive to 6G,  
for example large intelligent surfaces~\cite{han2019LIntegSurface} and holographic beamforming~\cite{chuang2020holographic}, 
are out of the  scope of this article.   
\section{Intelligence of Future Networks and  Autonomy}
\label{sec:AI}
In addition to  communication,  
6G will, exclusively,  offer  other services like computing, control, localization, sensing and cashing for very heterogeneous use cases in a highly complex and dynamic environment. 
Organizing resources of 6G networks  sustainably   while  meeting desired  KPIs   will  inevitably   need to be done autonomously. 
Leveraging AI advances is a must in achieving networks self organization. 
Rendering to  the huge gain observed from incorporating  AI techniques  in optimizing wireless networks, research community has been giving them more and more attention~\cite{zhang2019deep,mao2018deep, jiang2016machine}.
Through learning and prediction,   AI will be an efficient solution for achieving the convergence in  managing and allocating 6G network resources for various services~\cite{zhang20196G}. 
Softwarization~\cite{afolabi2018sliceSoftwariz}, cloudization~\cite{checko2014cloud}, virtualization~\cite{mijumbi2016management}, and slicing~\cite{zhang2017network} which are  main techniques for  5G network orchestration,
will be also  
expected to be important characteristics of 6G  autonomous network, AI enabled however.
AI algorithms will, in addition, be implemented at   network edge, for example at the smart wearable  {devices}, enabling collective intelligence and distributed autonomy.

 When it comes to optimizing  resources (time, spectrum, space, beam, mode, power, and code)  of multi-level and multi-dimensional networks with massiveness of connections,  
AI does not just outperform legacy techniques,  it however  emerges as a   main management technique for the following reasons:    
\begin{itemize}
\item Conventional design methods which are built based on  mathematical and statistical models require perfect system characteristic knowledge  which can not always be accessible. 
\item Traditional methods based solutions are not always optimal as their corresponding decisions are taking only according  to current input and no account is taken for  future information. 
\item Due to  {the} complexity of a wireless network problem,   mathematical solution{s} may not be feasible. 
\end{itemize}
Rending to their high ability  in learning, predicting patterns, and taking decisions accordingly, AI techniques, on the other hand, lead to enhancing wireless network performance in situations where legacy techniques fail to~\cite{adem2017jamming,Benfaid2021thesis, benfaid2021adaptsky,Badi2021SM}. 

\begin{figure*}
     \centering
     \begin{subfigure}{0.3\textwidth}
         \centering
         \includegraphics[width=\textwidth]{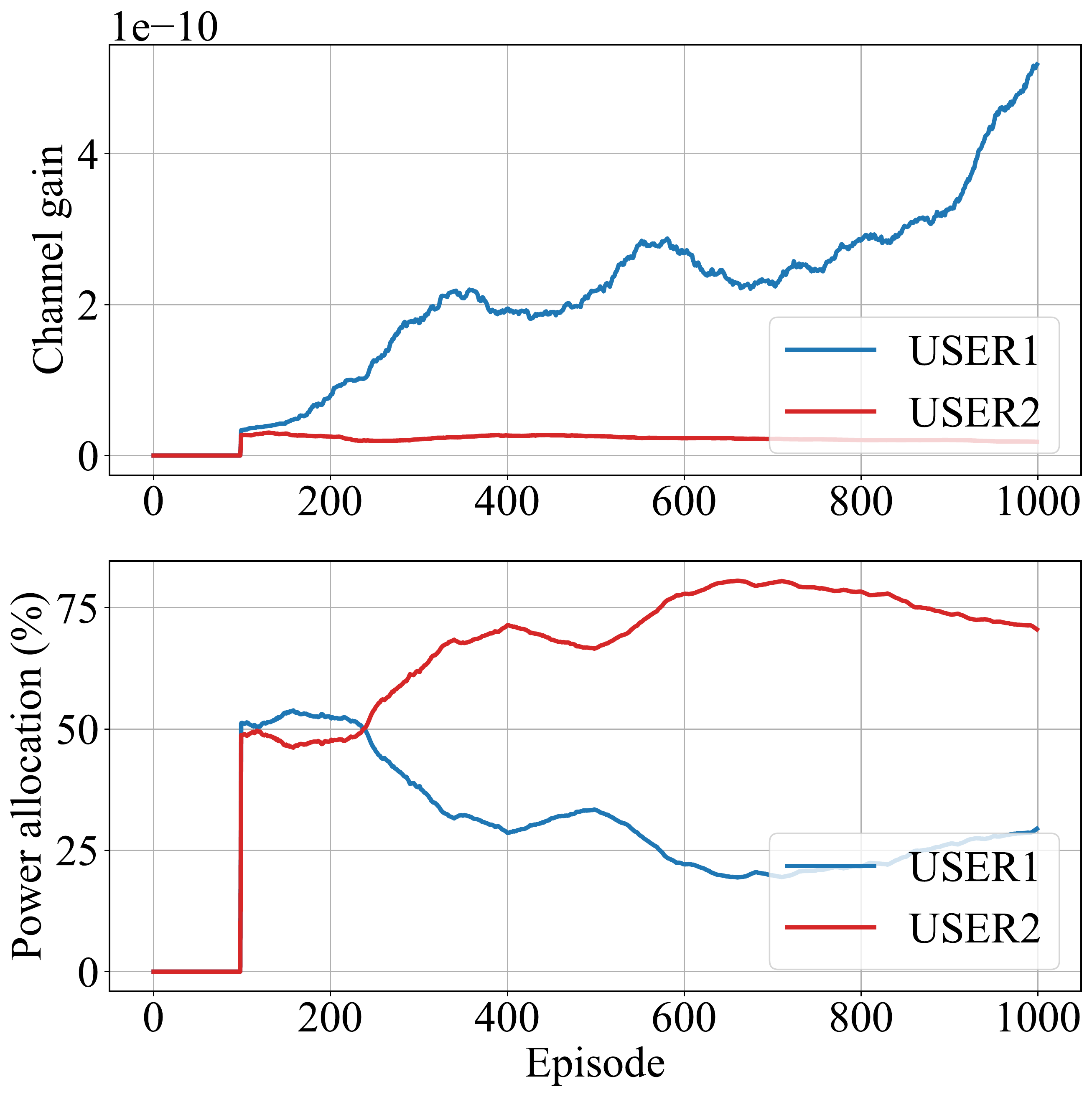}
         \caption{}
         \label{fig: LoS-Cluster_1}
     \end{subfigure}
     \hfill
     \begin{subfigure}{0.3\textwidth}
         \centering
         \includegraphics[width=\textwidth]{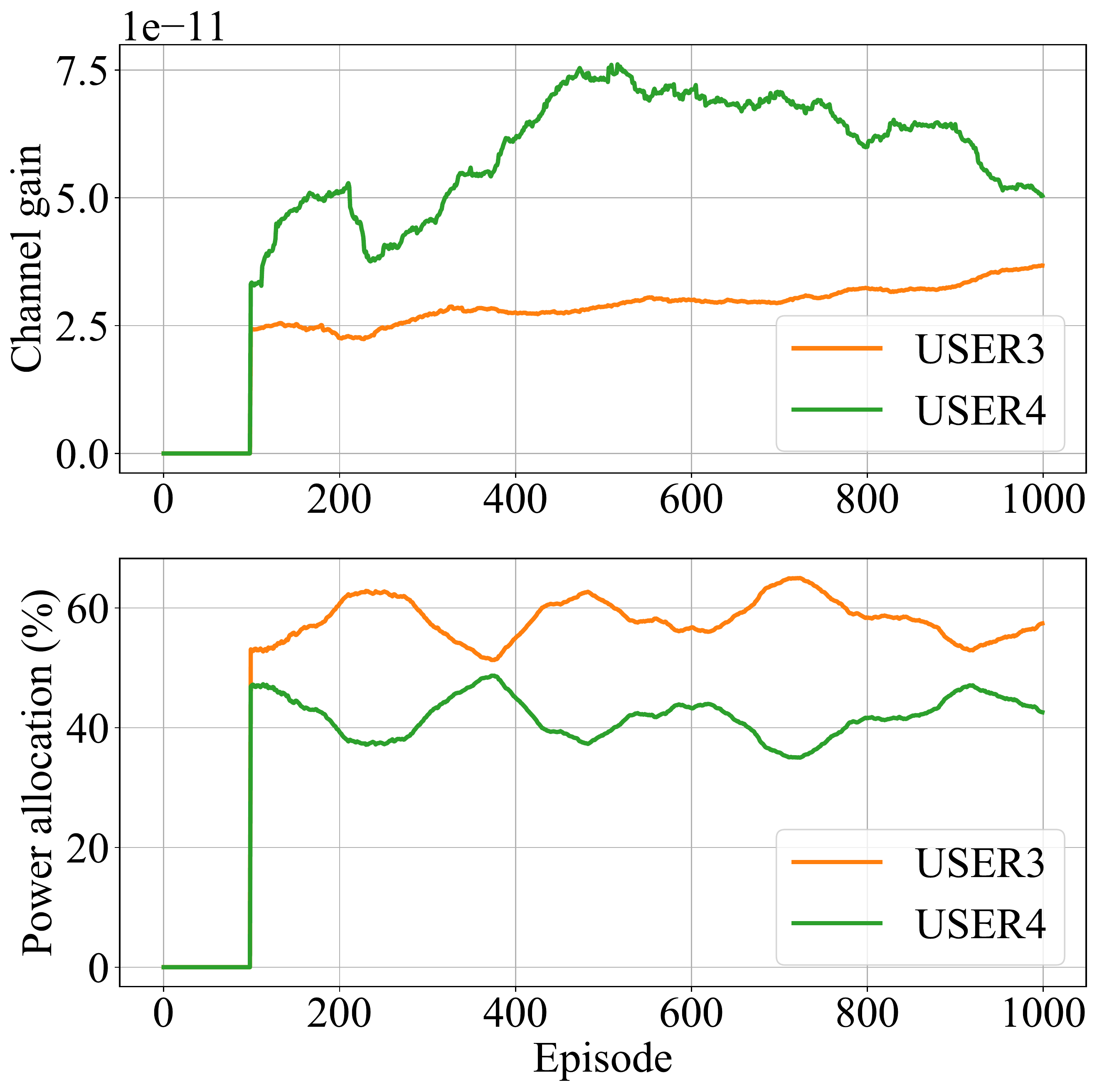}
         \caption{}
         \label{fig: LoS-Cluster_2}
     \end{subfigure}
     \hfill
     \begin{subfigure}{0.3\textwidth}
         \centering
         \includegraphics[width=\textwidth]{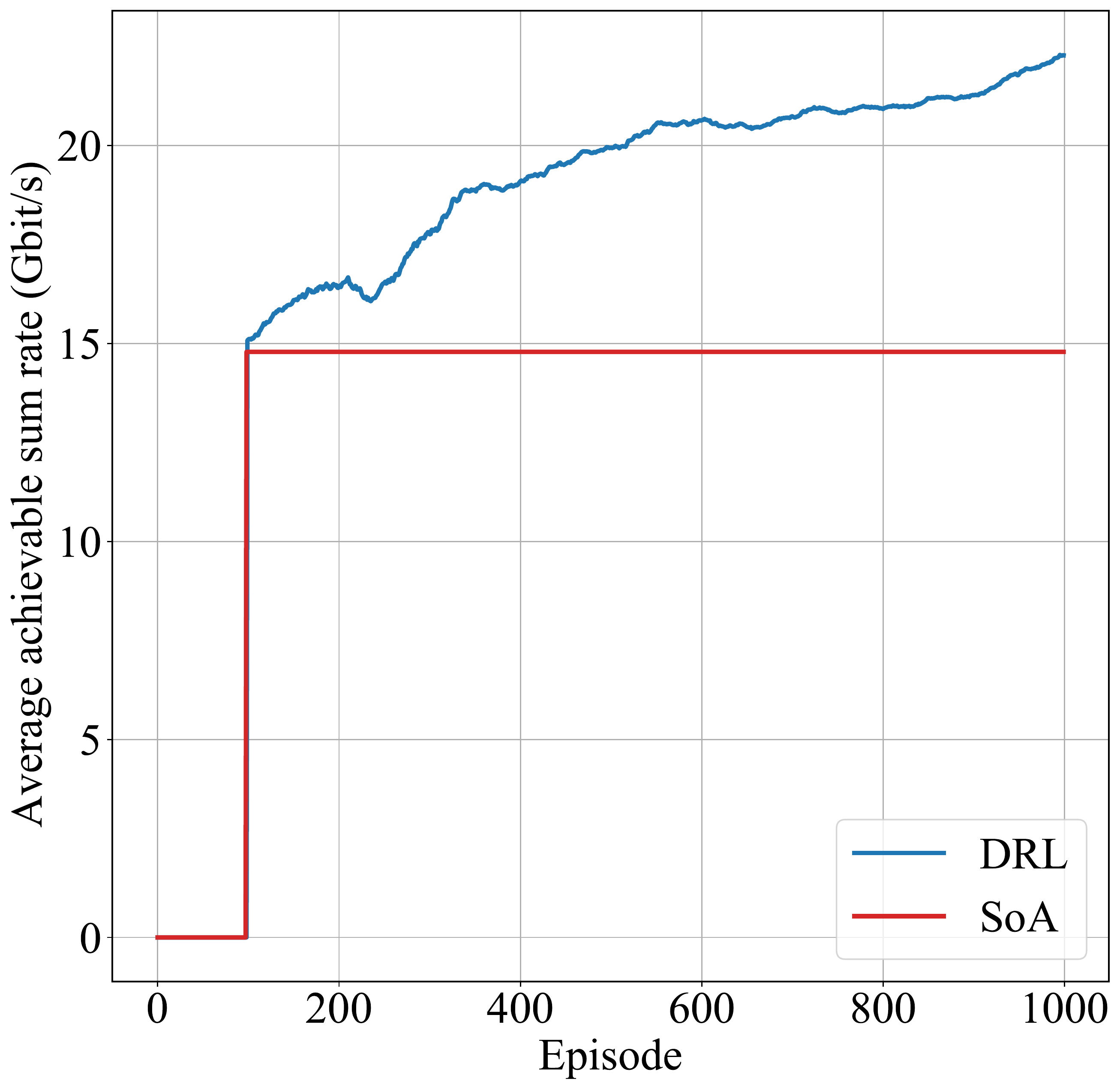}
         \caption{}
         \label{fig: LoS-Clusters-Sum-Rate}
     \end{subfigure}
        \caption{DRL decision and performance moving averages without forcing NOMA order (a) \& (b) cluster 1 \& 2 channel gain and power allocation respectively, and (c) total average achievable sum rate~\cite{Benfaid2021thesis}.}
        \label{fig: AdaptSky Clusters}
\end{figure*}

   Managing a network that integrates mmWave, NOMA, and UAV technologies such that  the  power and beam  allocation, and UAV placement are jointly optimized, for example,  is definitely a complicated task and  traditional methods fall short to handle~\cite{Benfaid2021thesis, benfaid2021adaptsky}.  
 Using novel-AI solutions, we   accomplished  in~\cite{Benfaid2021thesis,benfaid2021adaptsky} to not only solve the problem without the need {of} disjointing it into sub (unrealistic) ones but also  outperform some other existing heuristics derived  by legacy methods e.g.~\cite{Chen2019}. We, more specifically, proposed a  deep reinforcement learning (DRL)~\cite{mnih2013playing} based framework  that simultaneously places a  UAV in a 3D space  and allocates NOMA  power among users associated in clusters such that average sum of users\textquotesingle~data rate is maximized and certain fairness criteria is met. 
In this article, we present some of the simulation results as the UAV is being trained to learn the optimal decision while serving a total of four users, where each two are  associated with a certain cluster. The $ith$ user will be denoted by USERi and a decision interval will be referred to as an episode. 
More details about the framework and simulation setup are available in~\cite{Benfaid2021thesis,benfaid2021adaptsky}.
Fig. \ref{fig: AdaptSky Clusters} shows 100-episode moving of power allocation, channel gain, and  average sum rate of our proposed framework, denoted by DRL, as they vary over episodes.  In Fig. \ref{fig: AdaptSky Clusters}(c) we include the average sum rate achieved by the state of art (SoA) suggested in~\cite{Chen2019} and denoted in the figure by SoA. 
 We can notice that the DRL average achievable sum rate converges to $23.5$ Gbit/s which represents a $57\%$ improvement compared to $15$ Gbit/s given by SoA. 
  To be able to determine a  mathematical solution for the optimization problem,  the authors in~\cite{Chen2019}  ended up restricting  the UAV  placement into  a 2D plane. In contrary, our DRL framework allows  3D UAV mobility without making any confines. Hence, the gain in performance our AI based technique offers is intuitive. 
  Furthermore, we can notice that DRL managed to recognize the NOMA power allocation  order,  by allocating the far user (the one with the poor channel condition as depicted in the top of  Fig. \ref{fig: AdaptSky Clusters}(a) and (b)) more power (as exhibited in the bottom of Fig. \ref{fig: AdaptSky Clusters}(a) and (b)) than the near one without even imposing that in the algorithm. The resultant choice of order is crucial for the work of SIC, as discussed earlier,
  and is the best  to get the most out of NOMA. The  power allocation related decision was made intelligently  autonomously by the UAV through the judicious algorithm design with the goal of  improving  total sum rate and maintaining a certain fairness level. 
The presented findings 
do not only show how compelling  AI schemes are in enhancing performance, but also their abilities in  achieving autonomy and hence realizing  future networks vision.

\section{Future Network Deployment Challenges and Solutions}
\label{sec:chalg} 
\subsection{Complexity} 
 The superior performance of 6G and the ability to achieve full autonomous networks requires high computation capabilities, which will be a challenging task even for most powerful computers.

Despite its extremely wide spectrum, mmWave and THz technologies, for example, require a highly directional transmission which incur a substantial overhead. Searching in a very large angular space is essential for achieving high beamforming gain and thus, reducing the initial access  performance. The process of  aligning    beams between two communicating entities 
leads to communication and computational overhead. 
 The good news, however, is that ML  and DL-based initial access methods
have the potentials in reducing such  overhead~\cite{Mazin2018Accelerating,Myers2020Deep}. 
Moreover, the immense power of quantum computers in speeding up ML algorithms can be investigated not only for beam alignments, but also for some other services such as  users tracking~\cite{syed2019Quantum}. 

 NOMA, even though it enjoys  features that are important for ensuring meeting number of 6G objectives, 
  is still limited by 
 the high computation complexity incurred for SIC  estimation, especially, as stated earlier, if number of users is high. Moreover,  clustering and  allocating power for mobile users in mmWave-NOMA demand high computational power,   as they need to be updated in real time. Emerging  AI techniques can overcome some of these limitations~\cite{cui2018application}. 
 
With the ever-increasing demand for high capacity, implementing multi-user multiple input multiple output (MU-MIMO)  will be impractical. The utilized detection algorithms suffer from high complexity which increases exponentially with the number of users and data rate.~\cite{kim2019leveraging} is one  of the few works
that proposed a solution for MU-MIMO complexity problems and showed its effectiveness by implementing it on a real quantum computer.
With very low bit error rate  considering different modulation schemes and channel conditions,  the authors of~\cite{kim2019leveraging} managed to achieve MU-MIMO signal detection, just  {in order of microseconds}, for a number of users and data rates that are unfeasible to handle in  classical computers. 

The necessity for high computational power and parallelism suggests that quantum machine learning  can be a crucial candidate in 6G networks and beyond. 
\subsection{Security}
A number of applications including autonomous system such as healthcare, robotic, and vehicular communications demand a very-high level of security. Classical cryptographic methods would fail short with the emerging quantum technology~\cite{chen2016classCrypRept}. 
    Through some of its attributes including anonymization, decentralization, and untraceability, blockchain technology, however, can come as a solution for security vulnerability issues~\cite{henry2018blockchain}. 
    UAVs for example are emerging not just as aerial BSs, but also to provide a variety of other applications such as logistics, surveillance, disaster management, and rescue operations. Securing such UAV networks,  which will eventually be transformed into  Internet of UAVs, will inevitably require leveraging of  blockchain technology.
    Addressing   security issues  for which the unique 6G networks may be vulnerable to, will certainly  require a lot of attention from the  research community.

\section{Conclusion}
\label{sec:conc}
In this article we outlined the 6G vision and requirements. We presented some of 6G enabling technology. We also emphasized on  the need of 6G network autonomy and hence intelligence. In addition, considering some of our numerical analysis, we  emphasized on 
the necessity of integrating  AI techniques in the  resources management of  future networks. Furthermore, the article
presents some open research problems and challenges in realizing future networks objectives and trends,  and discusses the potentials  of some emerging  technologies in solving them.
\bibliographystyle{IEEEtran}
\bibliography{2References}

\end{document}